\documentclass[conference]{IEEEtran}
\IEEEoverridecommandlockouts
\usepackage{cite}
\usepackage{url}

\usepackage{amsmath,amssymb,amsfonts}
\usepackage{algorithmic}
\usepackage{graphicx}
\usepackage{multirow}
\usepackage{textcomp}
\usepackage{xcolor}
\def\BibTeX{{\rm B\kern-.05em{\sc i\kern-.025em b}\kern-.08em
    T\kern-.1667em\lower.7ex\hbox{E}\kern-.125emX}}
\begin{document}

\title{An analysis of large speech models-based representations for speech emotion recognition}

\author{\IEEEauthorblockN{Adrian Bogdan STÂNEA, Vlad STRILEȚCHI,  Cosmin STRILEȚCHI, Adriana STAN}
\IEEEauthorblockA{\textit{Communications Department} \\
\textit{Technical University of Cluj-Napoca, Romania}\\
\{adrianstanea1,vlad.striletchi\}@gmail.com\\
\{cosmin.striletchi, adriana.stan\}@com.utcluj.ro}
}

\IEEEoverridecommandlockouts
\IEEEpubid{\makebox[\columnwidth]{979-8-3503-2797-7/23/\$31.00 \copyright2023 IEEE\hfill} \hspace{\columnsep}\makebox[\columnwidth]{ }}

\maketitle

\IEEEpubidadjcol

\begin{abstract}

Large speech models-derived features have recently shown increased performance over signal-based features across multiple downstream tasks, even when the networks are not finetuned towards the target task. 
In this paper we show the results of an analysis of several signal- and neural models-derived features for speech emotion recognition. We use pretrained models and explore their inherent potential abstractions of emotions. Simple classification methods are used so as to not interfere or add knowledge to the task. We show that, even without finetuning, some of these large neural speech models' representations can enclose information that enables performances close to, and even beyond state-of-the-art results across six standard speech emotion recognition datasets. 
\end{abstract}


\section{Introduction}

Speech emotion recognition (SER) can provide an important additional information channel in various speech-enabled applications. As opposed to the linguistic contents which can be, to some extent, controlled by the speaker, the emotions lie deeper within the cognitive processes and require more active effort on behalf of the person. However, similar to speech variability, emotion realisation variability can add a higher complexity to their classification task. Also, in the absence of the visual channel, spoken emotions can be easily mistaken by the listeners. 

In terms of classification methods and architectures, the literature shows a wide interest for SER, with studies ranging from simple signal-based analyses to more recent deep architectures. In most of these studies, purposely built sets of features are used. 

Another problem with SER lies within the lack of extended spoken corpora with natural emotion elicitation. The common evaluation benchmarks for SER are a set of acted speech datasets, in which professional or naive speakers were instructed to perform a pre-defined or spontaneous spoken interaction in a desired emotion. 

In this paper we aim to perform an analysis over the use of various signal features, including large speech models- and self supervised learning-based (SSL) models-derived embeddings, in an effort to accurately classify the emotions present in 6 different datasets. We look into 9 sets of features and use simple classification methods to determine the intrinsic abstract representations of emotion that they may contain.

The paper is organised as follows: Section~\ref{sec:rel} gives a brief overview of the state-of-the-art on speech emotion recognition. Section~\ref{sec:meth} introduces the methodology and used speech datasets. Section~\ref{sec:results} presents and discusses the numeric results, while conclusions are drawn in Section~\ref{sec:conc}.

\section{Related work}
\label{sec:rel}

Emotion recognition has long been a topic of interest for the speech research community~\cite{el2011survey}. Initial studies focused on simple signal-based measures, such as Mel Frequency Cepstral Coefficients, Linear Predictive Coding Coefficients or the Teager Operator.  
With the advent of deep neural networks, the focus shifted towards large architectures which may be better at disentangling the emotion representation within the speech signal. Some of the recent studies attaining state-of-the-art results are focused on either recurrent or convolutional neural networks \cite{sarma18_interspeech,gmtcn,cpac,ye2023temporal}. For example, \cite{sarma18_interspeech} compares high-dimensional MFCC input  with frequency-domain and time-domain learning filters within the network. The main architecture is based on a time-delay neural network (TDNN). Ye~\textit{et al.} ~\cite{gmtcn} use a Gated Multi-scale Temporal Convolutional Network, which aims to create a unique component for learning emotional causality representation using a multi-scale receptive field that captures the temporal dynamics of emotions.  Additionally, the network incorporates skip connections to merge high-level features from different gated convolution blocks, enabling it to capture nuanced changes in emotion within human speech. GM-TCNet takes Mel-frequency cepstral coefficients as inputs and utilises the gated temporal convolutional module to generate the high-level features. Wen et al.~\cite{cpac} introduce an architecture based on capsule nets and transfer learning. It also deals with cross-corpus evaluation, for which it uses an adversarial module. 

Currently, the best performing network across several SER datasets is the one of~\cite{ye2023temporal}. The networks consists of so called temporally-aware blocks (TABs) which processes cropped 4 second-length segments from each utterance in a forward and reverse time flow. The TABs provide intermediate representations at different time-scales through dilated convolutions. The intermediate representations are dynamically fused together and passed through a final fully connected layer to output the final emotion class probabilities. 


Within this study, we also include the use of large pretrained speech models-derived features. By making use of vast amounts of speech data within their training step, these models are presumed to have learned abstract representations of speech, which are not easily noticeable or detectable with standard signal processing procedures. As a result, this type of features have been included in multiple other studies, such as~\cite{pepino2021emotion}, where the wav2vec features are passed through two dense layers, ReLU activation and dropout. The results are reported on two datasets, IEMOCAP~\cite{iemocap} and RAVDESS~\cite{ravdess}, and surpass the state-of-the-art results.

The main contribution of our work is linked to evaluating several large speech models trained on non-SER prediction tasks and the features they produce, as well as standard signal-based features (i.e. Mel frequency cepstral coefficients and Mel spectrogram) for SER classification. The features are passed through shallow and simple algorithms such that they do not learn additional high-level features in the process, but rather exploit the abstract or deterministic representations fed as input.

\section{Methodology}
\label{sec:meth}
To determine the extent to which very deep neural networks may perform an abstraction task that involves the representation of emotion, we select six standard SER datasets, extract nine feature sets, and use simple classification algorithms for evaluation. 

\subsection{Speech datasets}


We perform our analysis on a set of six widely adopted SER datasets: CREMA-D~\cite{cremad}, EMOVO~\cite{emovo}, EMODB~\cite{emodb}, IEMOCAP~\cite{iemocap}, RAVDESS~\cite{ravdess}, and SAVEE~\cite{savee}.
An overview of the these datasets is shown in Table~\ref{tbl:datasets}.\footnote{ For IEMOCAP we used only a balanced subset of the data containing around 5000 samples, and the excited emotion was merged into the happy category.} From Table~\ref{tbl:datasets} we can notice that not all emotions are present in all datasets, and in some of them the emotion categories may be merged or split according to the dataset designer's choice. This means that, aside from the complex task of speech emotion recognition, the lack of consensus across research groups with regards to the number and realisation of emotions in speech posses additional problems. Some other important facts to be noticed about the datasets is their different duration, as well as their multilingualism: IEMOCAP, RAVDESS, CREMA-D and SAVEE are English datasets, EMOVO is Italian, and EMODB is German. These axes of variation enable us to understand how the different speech features behave in a closer to real-life scenario, and if any of these features are in some way correlated to any of the axes. 


\begin{table*}[th!]
\centering
\caption{SER datasets used for evaluation. The M/F column refers to the male/female split amongst the speakers.}
\label{tbl:datasets}
\begin{tabular}{|lllcccc|}
\hline
\textbf{Dataset} & \textbf{Language} & \textbf{Emotions} & \textbf{Samples} & \textbf{Duration}  & \textbf{Speakers}  & \textbf{M/F}\\ \hline
\textbf{CREMA-D}~\cite{cremad} & Chinese & angry, disgusted, fearful, happy, neutral, sad & 7442 & 5h15'24 & 91 & 48/43\\ \hline
\textbf{EMOVO}~\cite{emovo} & Italian & angry, disgusted, fearful, happy, neutral, sad, surprised & 588 & 30' & 6 & 3/3\\ \hline
\textbf{EMODB}~\cite{emodb} & German & angry, disgusted, fearful, happy, neutral, sad, bored & 535 & 24' & 10 & 5/5\\ \hline
\textbf{IEMOCAP}~\cite{iemocap} & English & angry, happy, neutral, sad & 5531 & 11h 37' & 10 & 5/5\\ \hline
\textbf{RAVDESS}~\cite{ravdess} & English &  angry, disgusted, fearful, happy,       sad, surprised, calm & 1440 & 1h 28' & 24 & 12/12\\ \hline
\textbf{SAVEE}~\cite{savee} & English & angry, disgusted, fearful, happy, neutral, sad, surprised & 480 & 30' & 4 & 4/0\\
\hline
\end{tabular}
\end{table*}

\subsection{Speech Features and Pretrained Models}

\subsubsection{Signal features}

In terms of simple signal-based features we use two common representations within the speech research community:

\begin{itemize}
    \item{The \textbf{Mel Spectrogram} - is a simple auditory perception-based frequency shift of the speech spectrum. The Mel Spectrograms were extracted using the Librosa\footnote{\url{https://librosa.org/}} Python library. Each spectrogram was built using 80 filter banks to process the raw signal at a 22.05 kHz sampling rate. A Hanning window was used to weight the analysis window of length 2048, with a hop length of 512. The Mel Spectrograms were averaged across the time domain. }
    
    \item {\textbf{Mel Frequency Cepstral Coefficients (MFCC)} - are designed to mimic the human auditory system's perception of sound by emphasising relevant spectral information and reducing the impact of irrelevant noise and other audio variations. The cepstrum is a homomorphic transform based on the logarithmic representation of the speech spectrum. The Mel scale transform adds information with respect to the auditory perception of spectral components. The MFCCs were extracted using the Librosa library, using the same Hanning weighting and window length of 2048 with a hop length of 512. In this manner, 39 coefficients were extracted at each step and their mean value across time was computed for each individual signal.} 
\end{itemize}

\subsubsection{Speaker and language embedding networks}



For the current analysis we used a set of pretrained models available in the NeMo framework~\cite{nemo}.\footnote{\url{https://github.com/NVIDIA/NeMo}} NeMo is a large library of deep models released by NVIDIA which can be easily used within this framework. We selected the following pretrained models:

\begin{itemize}
    \item {\textbf{SpeakerNet}~\cite{koluguri2020speakernet} - uses 1D depth-wise separable convolutions, and x-vector based pooling layers. The architecture is lightweight and maps variable length utterances into a fixed length embedding;}
    \item {\textbf{TitaNet}~\cite{titanet} - uses the same 1D depth-wise separable convolutions, but combined with squeeze-and-excitation layers. A similar pooling layer is used to obtain a fixed length representation;}
    \item {\textbf{Ecapa TDNN} - is similar to the Speech Brain architecture and uses the ECAPA-TDNN structure~\cite{Desplanques_2020}. The difference is that, instead of the residual blocks, the NeMo implementation uses group convolution blocks of single dilation.}
    \item {\textbf{AmberNet} - it is very similar to the TitaNet architecture, yet its training objective was to discriminate between spoken languages. This network was selected in our study to verify if other speech-based tasks may also learn information about the speech emotions.}
\end{itemize}

\subsubsection{Self-Supervised Learning networks}

Self-supervised learning (SSL) networks are omnipresent in various speech-related applications with results going beyond the state-of-the-art across numerous tasks. Their main advantage is the fact that within their training procedure, they are guided towards finding an optimal compressed representation of speech. As long as the representation is not finetuned towards a target task, it can be assumed that it includes the different dimensions of speech variability, such as speaker, intonation, recording conditions, and possibly, even emotions. 

We, therefore, include in our study a set of three SSL pretrained models available in the S3PRL\footnote{\url{https://github.com/s3prl/s3prl}} framework. S3PRL is built as a wrapper over multiple speech models released by various research groups. This unifies the use of these models and makes it easier to test several representations in downstream tasks. We selected the following three models:

\begin{itemize}
    \item {\textbf{wav2vec}~\cite{wav2vec} - is one of the most commonly used self-supervised representations across various applications. Its initial task was that of automatic speech recognition. It uses a convolutional neural network (CNN) to directly learn representations from raw audio waveforms without requiring any explicit phonetic or linguistic annotations. The model utilises a contrastive objective function to encourage the encoder to learn to differentiate between positive and negative examples, enhancing the robustness and generalisation of the learned representations.}
    \item {\textbf{Decoar}~\cite{decoar} - is a semi-supervised model which uses a large amount of unlabelled audio data for an initial representation learning. The representation is based on filterbank features extracted from a context frame. The so called deep contextualised acoustic representations (DeCoAR) are then fine-tuned towards an automatic speech recognition system.}
    \item {\textbf{HuBERT}~\cite{hubert} - was released as a speech-based version of the BERT text models.  It utilises a combination of contrastive learning and the model's transformer-based architecture to extract discriminative and context-rich features from raw audio waveforms. HuBERT has demonstrated impressive performance in pretraining speech representations and has been used as a backbone for various downstream speech processing tasks, such as automatic speech recognition (ASR) and speaker verification.}
\end{itemize}

No finetuning of any of the deep models was performed, and the official releases for them were used. All networks take raw waveforms as input. 

\subsection{Data Pre-processing and SER Prediction Models}

The main purpose of this study is to perform an analysis over various speech representations that may inherently incorporate high-level information regarding the emotions elicited in the utterances. And we did not aim at surpassing the state-of-the-art results in speech emotion recognition. Therefore, we only used lightweight classification algorithms trained on the speech representations described in the previous subsection. 

We selected two classification methods: a support vector machine-based classifier (SVC) and a four layer deep feed forward neural network (DNN): 

\begin{itemize}
    \item The used SVC is the one implemented in Scikit-Learn\footnote{\url{https://scikit-learn.org/stable/}} and had a the regularisation parameter equal to 10 and the gamma parameter was set to 0.001;
    \item The DNN consists of a series of layers, specifically designed for the task of emotion classification. The input layer accepts the feature tensor for processing. Subsequently, a set of two hidden layers incrementally transform the input into a more abstract representation by halving the output size with each layer, which results in the third layer having an output size that is 8 times smaller than the initial feature vector. Complementary to this transformation, a batch normalisation layer and a Rectified Linear Unit (ReLU) activation function are utilised alongside a dropout layer to prevent overfitting. The DNN was implemented in Pytorch.\footnote{\url{https://pytorch.org/}}     
\end{itemize}

Also, because our task is not to temporally locate the emotions within a spoken utterance, we used time-domain averaged representations. As a result, the only change across the classification algorithms is the number of input features. 

\section{Results}
\label{sec:results}

\begin{table*}[h!]
\centering
\label{tbl:results}
\caption{UAR[\%] and WAR[\%] results for all datasets and feature sets. The last column averages the UAR and WAR measures across the datasets. Best results are marked in boldface.}
\begin{tabular}{|l|c|cc|cc|cc|cc|cc|cc||cc|}
\hline
\multirow{3}{*}{\textbf{Feature set}} & \multirow{3}{*}{\textbf{Model}} & \multicolumn{12}{c|}{\textbf{Dataset}} & \multicolumn{2}{c|}{} \\ \cline{3-14}
&  & \multicolumn{2}{c}{\textbf{CREMA-D}} & \multicolumn{2}{c|}{\textbf{EMOVO}} & \multicolumn{2}{c|}{\textbf{EMODB}} &\multicolumn{2}{c|}{\textbf{IEMOCAP}} & \multicolumn{2}{c|}{\textbf{RAVDESS}} &\multicolumn{2}{c|}{\textbf{SAVEE}} & \multicolumn{2}{c|}{\textbf{Feature Mean}}\\ \cline{3-16}
& & UAR & WAR & UAR & WAR &UAR & WAR &UAR & WAR &UAR & WAR &UAR & WAR & UAR & WAR \\ \hline \hline

\multirow{2}{*}{\textbf{Mel Spec.}} & SVM & 37.25  & 37.68 & 35.22 & 34.70 & 41.66 & 48.22 & 35.84 & 41.24 & 21.22 & 21.39 & 38.65 & 43.54 & 34.97 & 37.80 \\
                                & DNN & 50.98 & 41.56 & 61.65 & 50.21 & 65.1 & 66.04 & 50.33 & 42.96 & 42.44 & 30.9 & 69.89 & 59.38 & 56.73 & 48.51 \\ \hline
                                
\multirow{2}{*}{\textbf{MFCCs}} & SVM & 47.83  & 47.86 & 66.98  & 67.19 & 67.94 & 70.47 & 59.12 & 58.51 & 61.96 & 61.39 & 68.45 & 70.62 & 62.05 & 62.67 \\
                      & DNN & 49.44 & 44.49 & 70.92 & 66.04 & 74.25 & 72.08 & 60.04 & 57.98 & 47.02 & 48.06 & 72.72 & 67.29 & 62.40 & 59.32\\ \hline
 \hline

\multirow{2}{*}{\textbf{SpeakerNet}} & SVM & 58.90  & 58.88 & 75.79 & 75.35 & 90.56 & 90.09 & 67.32 & 66.86 & 79.44 & 78.68 & 72.84 & 75.00 & 74.14 & 74.14 \\
                            & DNN & 57.09 & 57.22 & 84.84 & 83.96 & 95.82 & 95.21 & 67.27  & 66.27 & 82.95 & 82.78 & 79.88 & 80.83 & 77.97 & 77.71\\ \hline
                                        
\multirow{2}{*}{\textbf{TitaNet}}    & SVM & 65.41 & 65.36 & 85.58 & 85.04 & 94.99 & 94.58 & 69.78 & 69.39 & 80.97 & 80.69 & 83.11 & 85.21 & 79.97 & 80.05 \\
                            & DNN & 63.61 & 63.54 & \textbf{91.73} & \textbf{91.25} & \textbf{98.34} & \textbf{97.92} & 70.63 & 69.89  & 85.18 & 85.49 & \textbf{88.94} & \textbf{88.75} & \textbf{83.07} & \textbf{82.81}\\ \hline
                            
\multirow{2}{*}{\textbf{Ecapa TDNN}}  & SVM & 59.63  & 59.65  & 82.10 & 81.82 & 93.18 & 93.27 & 68.47 & 68.00 & 76.27 & 76.25 & 77.25 & 79.17 & 76.15 & 76.36 \\
                             & DNN & 60.04 & 60.04 & 88.79 & 88.54 & 97.29 & 97.08 & 68.9 & 68.09 & 81.42 & 81.74 & 84.57 & 84.38 & 80.17 & 79.98 \\ \hline

\multirow{2}{*}{\textbf{AmberNet}}   & SVM & 68.78  & 68.64 & 69.52 & 69.23 & 89.78 & 90.47 & 71.62 & 71.18 & 87.24 & 87.43 & 77.30 & 78.96 & 77.37 & 77.65 \\
                            & DNN & 70.53 & 70.2 & 80.16 & 79.58 & 94.94 & 94.38 & 71.92 & 70.92 & \textbf{89.41} & \textbf{89.65} & 85.36 & 85.21 & 82.05 & 81.66 \\ \hline    \hline                        
     
\multirow{2}{*}{\textbf{wav2vec}} & SVM & 67.77 & 67.63 & 71.29 & 71.43 & 93.53 & 93.27 & 70.29 & 69.50 & 73.64 & 73.89 & 67.33 & 69.38 & 73.97 & 74.18 \\
                         & DNN & 69.38 & 69.18 & 82.33 & 82.5 & 96.83 & 96.25 & 69.09 & 68.33  & 78.96 & 79.51 & 75.75 & 75.83 & 78.72 & 78.60 \\ \hline
                                        
\multirow{2}{*}{\textbf{Decoar}} & SVM & 67.61 & 67.43 & 65.83 & 65.98 & 86.27 & 87.29 & \textbf{71.93} & \textbf{71.42} & 70.73  & 71.11 & 57.11 & 60.21 & 69.91 & 70.57 \\
                        & DNN & 70.04 & 69.8 & 83.72 & 82.5 & 97.82 & 97.29 & 70.75 & 69.8  & 80.34 & 80.63 & 76.02 & 74.58 & 79.78 & 79.10\\ \hline

\multirow{2}{*}{\textbf{HuBERT}} & SVM & 70.72  & 70.46 & 31.15 & 30.44 & 89.70 & 90.09 & 71.60 & 70.91 & 80.97 & 80.69 & 57.50 & 61.88 & 66.94 & 67.41 \\
                                         & DNN & \textbf{72.62} & \textbf{72.50} & 39.14 & 37.92 & 96.13 & 95.63 & 71.43  & 70.28  & 76.18 & 75.97 & 69.21 & 69.79 & 70.78 & 70.35\\ \hline \hline

\textbf{State-of-the-Art} & - & 82.96 & N/A & 92.00 & 92.00 & 95.17  & 95.70  & 72.50 & 71.65 & 91.93 & 92.08  & 86.07  & 87.71 &  - & - \\ \hline

\end{tabular}
\end{table*}

The results of our analysis are shown in Table~\ref{tbl:results}. We also list the state-of-the-art results across the six datasets. These pertain to the following studies:  
for CREMA-D -- \cite{10049941}; IEMOCAP --\cite{lian20b_interspeech}; for EMOVO, EMODB, RAVDESS, and SAVEE -- \cite{ye2023temporal}. We need to note here that since our submission, Ye \textit{et al.}~\cite{ye2023temporal} have updated their reported results with lower performance on average. Yet we maintain their top reported results in order to show that the different embeddings still include a large amount of non-speaker related information. 

The performance measures are the unweighted average recall ($UAR$) and the weighted average recall ($WAR$):

\begin{equation}
UAR = \sum_{k}\frac{TP_k}{TP_k+FN_k}
\end{equation}

\begin{equation}
WAR = \sum_{k}\alpha_k\frac{TP_k}{TP_k+FN_k}
\end{equation}

\noindent where $k$ is the emotion class id, and $\alpha_k$ is the weighting coefficient for the respective class. $\alpha_k$ is computed as the fraction of samples which pertain to class $k$ within the complete set of test samples. $UAR$ is a special case of $WAR$ where all $\alpha_k$ are equal to $1/k$.
$TP$ counts the true positive samples of a class, and $FN$ refers to the false negative counts. The recall, as opposed to the accuracy, measures how many positive samples were correctly identified from the total number of positive examples, rather than a total count from all classes. In a multi-class classification task, the recall is computed for each individual class, and then averaged to obtain the task's recall measure. If the same weight is given to each class, irrespective of its representation within the test set, the recall is then \emph{unweighted}. The unweighted average recall in multi-class classification is equal to the accuracy. On the other hand, if the percentage of samples from a class within the test set is used to weight the class recall, then the final measure is called \emph{weighted average recall}. The recall measure has been widely adopted in the SER studies, and we adopt it here, as well.  
Also because there are no official splits of the datasets into training and testing subsets, we perform a 5-fold cross validation for each dataset and report the averaged results.

Some very important results arise from the table: except for the CREMA-D dataset, all our results are very close, and in some cases surpass the state-of-the-art results (last row of the table). For the EMODB and SAVEE datasets, the TitaNet features combined with the simple DNN algorithm show an average 2\% absolute increase in both the UAR and WAR measures. 
The best performing set of features across all the datasets is that extracted with the TitaNet architecture. 

As expected, simple signal-derived features (i.e. Mel spectrogram and MFCCs) are not as proficient as the DNN-derived ones. One interesting aspect is that the AmberNet architecture which was trained for language recognition is able to provide rich speech representations that include discriminative information about the speech emotion. Also, all speaker embedding networks have a better performance, as opposed to the SSL-features which seem to be less representative for the emotion prediction.  A side result of this analysis is again the fact that there representations, although trained to disentangle the speech variation axes, are not yet providing such decomposition power, similar to the results shown in~\cite{stan2022}.

\section{Conclusions}
\label{sec:conc}

In this paper, we performed an analysis over several speech feature sets to find the best representation for speech emotion recognition. These feature sets are either derived from signal processing, such as Mel spectrograms or MFCCs, while a separate set of features come from large pretrained speech models. Because we did not want to surpass the state-of-the-art results on SER, but rather to look into better emotion representations, the prediction algorithms in the evaluation were rather simple. An SVM classifier and a four layer feed forward network were used to predict the emotion classes across six SER datasets. 

Our evaluation showed that, at least for some of the datasets, these LSM-derived features already include relevant information for emotion recognition, although they were not trained for this particular objective. As a result, for example the embeddings from TitaNet and Decoar features on the EMODB, SAVEE, IEMOCAP datasets are very close or surpass the state-of-the-art results. We may argue that in this case, the use of more advanced prediction architectures can be convoluted, and that more attention should be paid to the speech representation. 

As future work, we plan to include the best feature sets resulted from this study into state-of-the-art networks purposely designed for SER. We assume that these features should simplify this complex task and attain better results. A separate task is that of finding more accurate speech datasets, either by sampling from the online resources and manually annotating the emotion information, or by accessing real data from relevant sources such as call-centres where this information may already be annotated by the operators.  

\section*{Acknowledgement}
The work presented in this paper was funded by the National Competitiveness Operational Programme, Romania, under project number SMIS 156387 - VOITA.


\bibliographystyle{IEEEtran}
\bibliography{main}

\end{document}